\documentclass{aastex52/aastex}
\usepackage{spr-astr-addons}    %% mimicing ASTR journal style
\begin{document}
%% Article title
%
\title{Properties of the ionized gas of circumnuclear star-forming regions in early type spirals}

%% Running heads
\shorttitle{The ionized gas of circumnuclear star-forming regions}
\shortauthors{D\'\i az et al.}

%% Author and Affilations
\author{\'Angeles I. D\'\i az, Guillermo F. H\"agele} 
\affil{Dpto de F\'{\i}sica Te\'orica, C-XI, Universidad Aut\'onoma de
Madrid, 28049 Madrid, Spain}
%\author{Guillermo F. H\"agele \altaffilmark{1}}
\and 
\author{Elena Terlevich \altaffilmark{1}, Roberto Terlevich \altaffilmark{1}}
%\author{Roberto Terlevich\altaffilmark{2}}
\affil{INAOE, Tonantzintla, Apdo. Postal 51, 72000 Puebla, M\'exico}
%\affil{}
%\email{} %% non-output

%% Alternate Affilations
%\altaffiltext{1}{Depto. de F\'\i sica Te\'orica, Universidad Aut\'onoma de Madrid, Spain}
%\altaffiltext{2}{Inistituto de Astrof\'\i sica, \'Optica y Electr\'onica, INAOE, M\'exico}
%\altaffiltext{3}{}
\altaffiltext{1}{Research Affiliate at IoA.}

%Abstract
\begin{abstract}
A study of cicumnuclear star-forming regions (CNSFRs) in several early type spirals has been made in order to investigate their main properties: stellar and gas kinematics, dynamical masses, ionising stellar masses, chemical abundances and other properties of the ionised gas. Both high resolution (R$ \sim $20000) and moderate resolution (R $ \sim $ 5000) have been used.

 In some cases these
regions, about 100 to 150\,pc in size, are seen to be composed of several
individual star clusters with sizes between 1.5 and 4.9\,pc estimated from
Hubble Space Telescope (HST) images.  Stellar and gas velocity dispersions
are found to differ by about 20 to 30\,km/s with the H$\beta$ emission lines
being narrower than both the stellar lines and the
[OIII]\,$\lambda$\,5007\,\AA lines. The twice ionized oxygen, on the
other hand, shows velocity dispersions comparable to those shown by stars.
We have applied the virial theorem to estimate dynamical masses of the
clusters, assuming that systems are gravitationally bounded and spherically
symmetric, and using previously measured sizes. The measured values of the 
stellar velocity dispersions yield dynamical masses of the order of 10$^7$ to
10$^8$ solar masses for the whole CNSFRs.

We obtain oxygen abundances which are comparable to those found in high metallicity disc HII regions from direct measurements of electron temperatures and consistent with solar values within the errors. The region with the highest oxygen abundance is R3+R4 in NGC~3504, 12+log(O/H) = 8.85, about 1.5 solar if the solar oxygen abundance is set at the value derived by \citet{AGS05}, 12+log(O/H)$_{\odot}$ = 8.66$\pm$0.05. The derived N/O ratios are in average larger than those found in high metallicity disc HII regions and they do not seem to follow the trend of N/O vs O/H which marks the secondary behaviour of nitrogen. On the other hand, the S/O ratios span a very narrow range between 0.6 and 0.8 of the solar value. As compared to high metallicity disc HII regions, CNSFR show values of the O$_{23}$ and the N2 parameters whose distributions are shifted to lower and higher values respectively, hence, even though their derived oxygen and sulphur abundances are similar, higher values would in principle be obtained for the CNSFR  if pure empirical methods were used to estimate abundances. CNSFR also show lower ionisation parameters than their disc counterparts, as derived from the [SII]/[SIII]. Their ionisation structure also seems to be different with CNSFR showing radiation field properties more similar to HII galaxies than to disc high metallicity HII regions.

\end{abstract}

%% Keywords
%\keywords{}

%%  Please use labels (\label, \ref) for section, figure, table, 
%%  equation  reference. Use \cite for bibliography references.
%
\section{Introduction}
\label{introduction}

The inner ($\sim$ 1Kpc) parts of some spiral galaxies show star formation complexes frequently arranged in an annular pattern around their nuclei. These complexes are sometimes called "hot spots" and we will refer to these as circumnuclear starforming regions (CNSFRs). Their sizes go from a few tens to a few hundreds of pc
\citep[e.g.][]{DATTSA00} and they seem to be made of an ensamble of HII regions
ionised by luminous compact stellar clusters whose sizes, as measured from
high spatial resolution HST images, are seen to be of only a few pc. 

The luminosities of CNSFRs are rather large with absolute visual magnitudes (M$_v$) between -12 and -17 and H$\alpha$ luminosities which are comparable to those shown by 30 Dor, the largest HII region in the LMC, and overlap with those shown by HII galaxies 
\citep[][and references therein]{MTM88,DATTSA00,HD06}. 
In the ultraviolet (UV), massive stars dominate the observed
circumnuclear emission even in the presence of an active nucleus
\cite{GD98,CGML02}.  

In many cases, CNSFR show emission line spectra similar to those of disc HII regions. However, they show a higher continuum from background stellar populations as expected from their circumnuclear location.  The analysis of these spectra gives us the oportunity to measure the gas abundances close to galactic nuclei which,  in the case of early type spirals, are expected to be amongst the highest metallicity regions.

The importance of an accurate determination of the abundances of high metallicity HII regions cannot be overestimated since they constitute most of the HII regions in early spiral galaxies (Sa to Sbc) and the inner regions of most late  type ones (Sc to Sd) \cite{D89} without  which our description of the metallicity distribution in galaxies cannot be  complete.  In particular, the effects of the choice of different calibrations  on the derivation of abundance gradients can be very important since any abundance  profile fit will be strongly biased towards data points at the ends of the  distribution. It should be kept in mind that abundance gradients are widely  used to constrain chemical evolution models, histories of star formation over galactic discs or galaxy formation scenarios. Also, the question of how high is the highest oxygen abundance in the gaseous phase of galaxies is still standing and extrapolation of known radial abundance gradients would point to CNSFR as the most probable sites for these high metallicities. 

Accurate measures of elemental abundances of high metallicity regions are also crucial to obtain reliable calibrations of empirical abundance estimators, widely used but poorly constrained, whose choice can severely bias results obtained for quantities of the highest relevance for the study of galactic evolution like the luminosity-metallicity (L-Z) relation for galaxies.

\section{Observations}
\label{Observations}
As part of the programme to study the properties of CNSFR we obtained two sets of data with the Intermediate dispersion Spectrograph and Imaging System  (ISIS) attached to the  4.2\,m William Herschel Telescope (WHT). The first set of observations consisted of high resolution blue and far-red long-slit spectra covering the emission lines of H$\beta$ and [OIII]  and the CaII triplet lines respectively. At the attained resolution of 0.4 and 0.7 \AA\ respectively, these data allowed the measurement of radial velocities and velocity dispersions of the gas and stars in the regions down to 25 km s$^{-1}$. The second set, also obtained with the same instrumentation had a lower resolution of 3.4 and 1.7 \AA\ in the blue and red spectral regions respectively and a wider total coverage from 3600 to 9650 \AA . These data were used to derive the physical conditions and the abundances of the emitting gas. 

\begin{figure}%[tb]
\plottwo{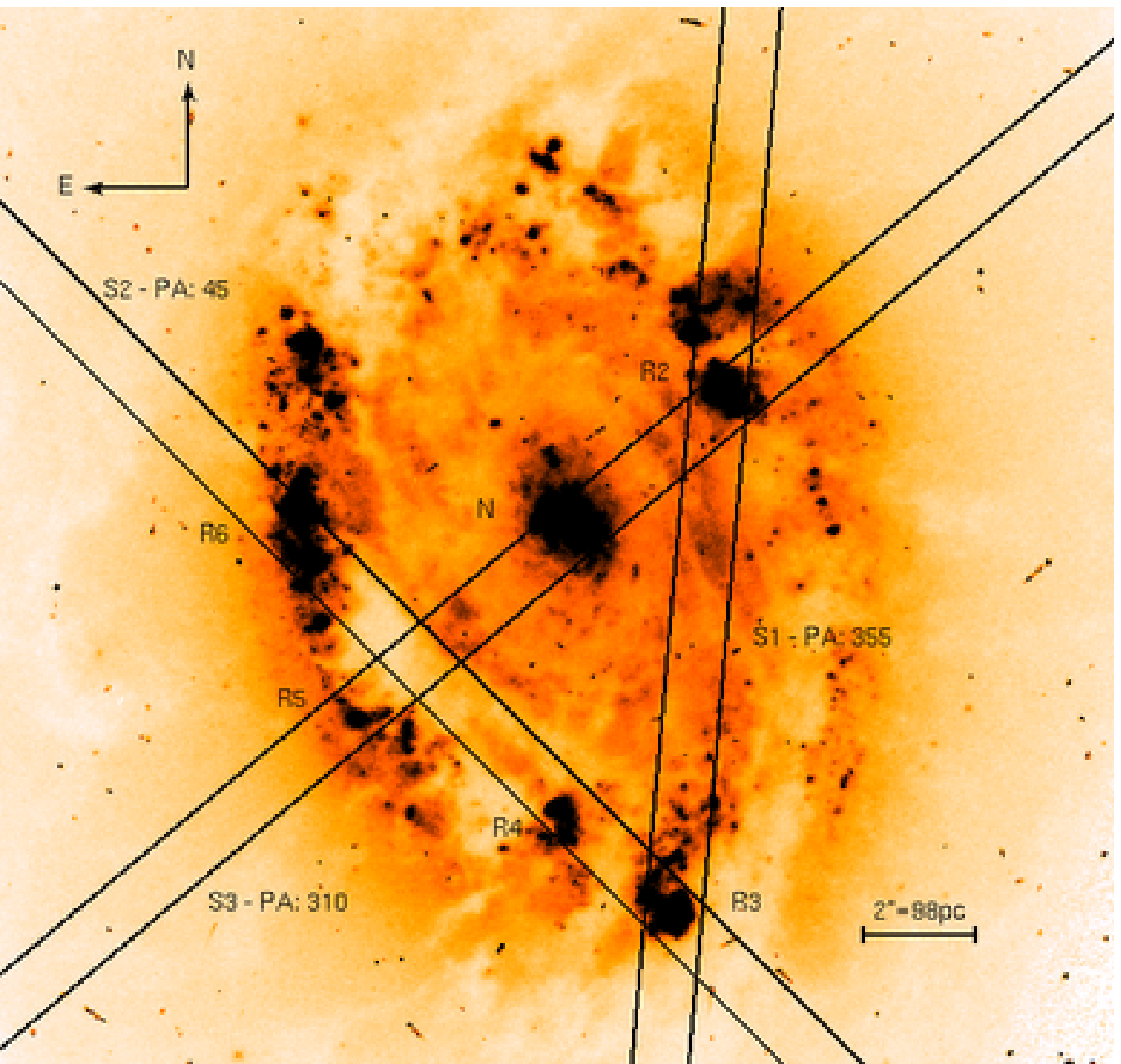}{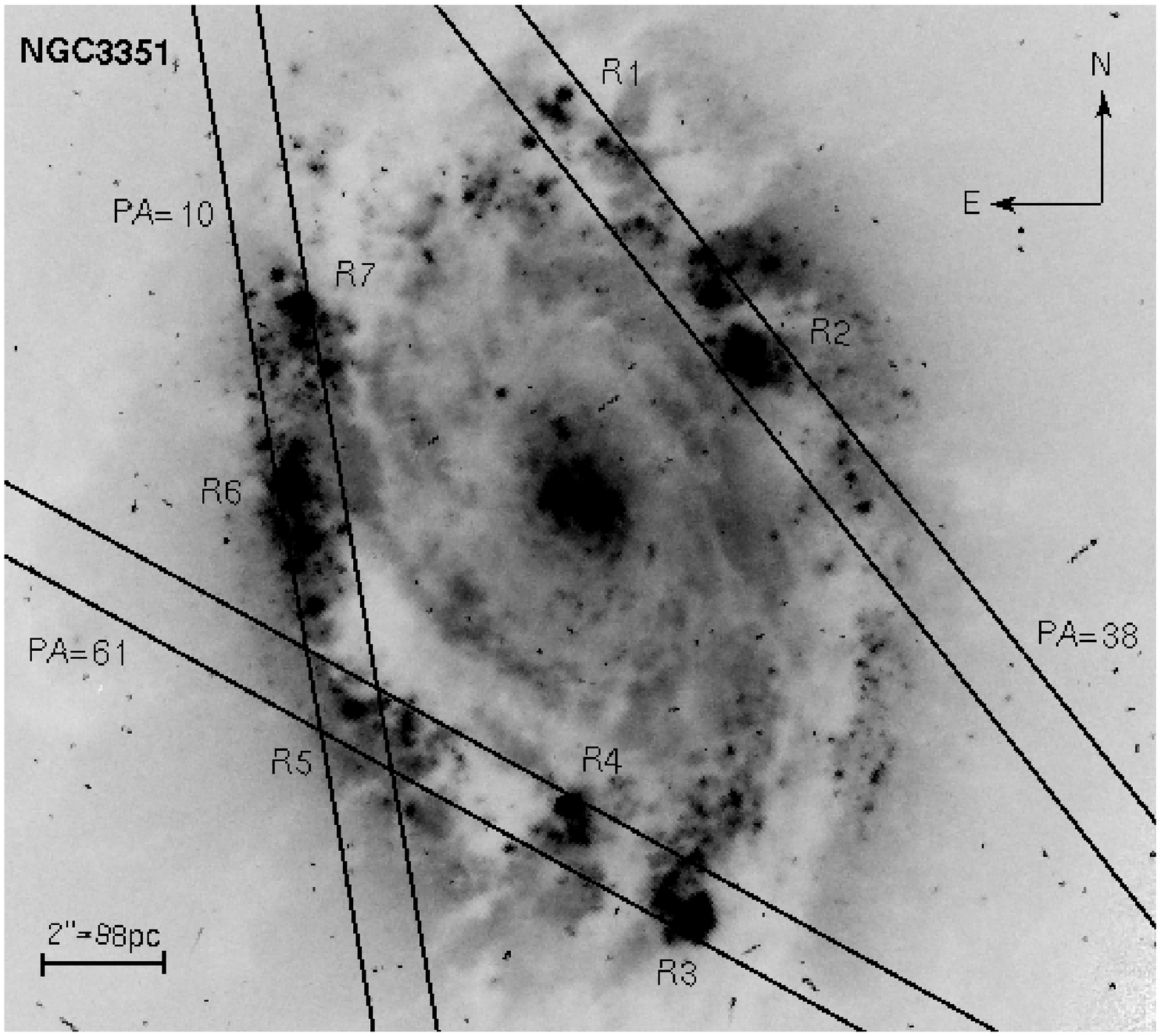}
\caption{F606W (wide V) image centred on NGC\,3351 obtained with the WFPC2
  camera  of the Hubble Space Telescope showing the WHT-ISIS slit positions overplotted for the kinematics set observations (left) and the abundance set observations (right). North is up and  East to the left } %% no full stop at the end of caption
\label{images}
\end{figure}

\begin{figure}[tb]
\plotone{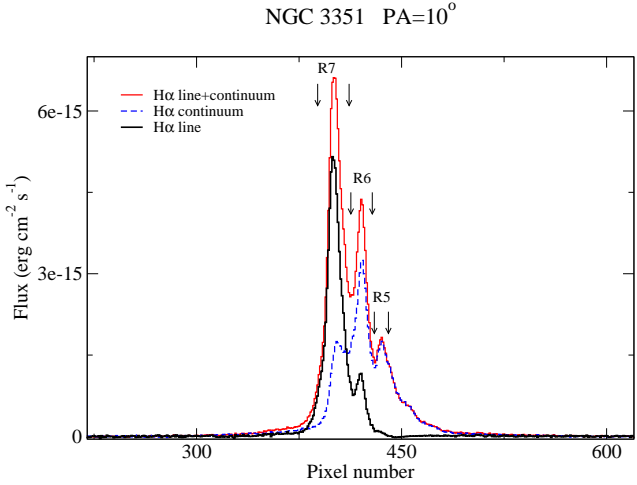}
\caption{Spatial light profiles along one of the slit position in NGC~3351. The thin solid (red) line shows the total H$\alpha$ emission, the thin broken (blue) line shows the continuum adjacent to H$\alpha$ and the thick solid line (black) shown the pure H$\alpha$ } %% no full stop at the end of caption
\label{profiles}
\end{figure}

Several CNSFR were observed in the spiral galaxies: NGC~2903, NGC~3351 and NGC~3504. Archive images obtained with the Wide Field and Planetary Camera 2 and the NICMOS Camera 3 both on-board the HST were also downloaded in order to complement the study.
Figure \ref{images} shows the optical image of NGC~3351 with the slit positions used for the two sets of observations. Figure \ref{profiles} shows the spatial distribution of the H$\beta$ and continuum emission along one of the slit positions observed. As can be seen from these profiles, some clusters are dominated by continuum emission while in other cases gaseous line emission is clearly important.

%% Figure 
%
%\begin{figure*}
  %  \centering
    %\includegraphics[width=0.48\textwidth]{NGC3351_sl_kin.eps}
%\hspace{0.2cm}
%\includegraphics[width=0.48\textwidth]{NGC3351_sl_ab.eps}
%\caption[]{Left: F606W (wide V) image centred on NGC\,3351 obtained with the WFPC2
  %camera (PC1) of the Hubble Space Telescope. Right: HST-NICMOS image obtained through the F160W filter. For both images the orientation is north up, east to the left. The location and P.A. of the WHT-ISIS slit positions, together with identifications of the CNSFRs extracted, are marked.}
%\label{images}
%\end{figure*}

\section{Summary of kinematical results}
\label{Kinematics}
\begin{figure}[tb]
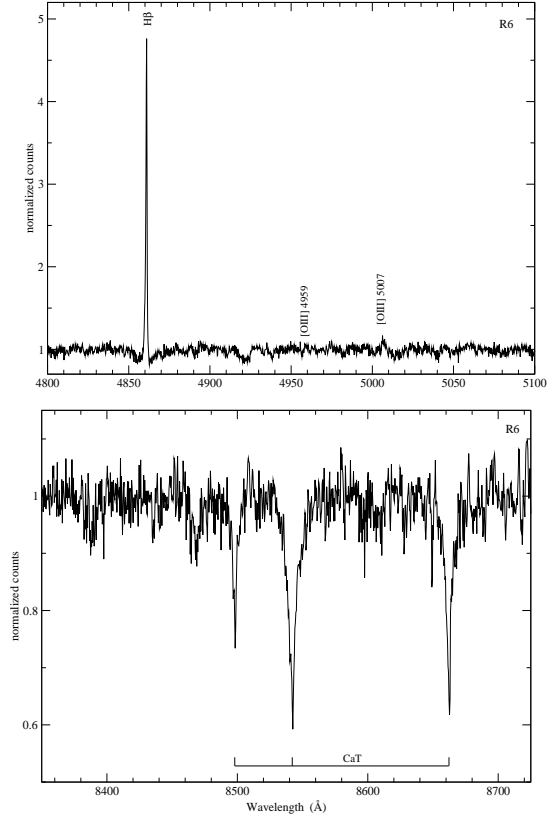

%\centering
\hspace{0.2cm}
\includegraphics[width=.40\textwidth,height=.31\textwidth,angle=0]{1dn_R6-s2-b-rest.eps}\\
\vspace{0.2cm}
\includegraphics[width=.40\textwidth,height=.31\textwidth,angle=0]{1dn_R6-s2-r-rest.eps}
\caption{Blue (upper panel) and red (lower panel) rest frame normalised
  spectra of R6. Notice the absence of conspicuous emission lines in the  
  red spectral range for this region.} 
\label{spectra-kin}
\end{figure}

Gas velocity dispersions were measured  by performing Gaussian fits to the H$\beta$\,$\lambda$\,4861\,\AA\ and [OIII] $\lambda$ 5007 \AA\ lines on the high dispersion spectra (Figure \ref{spectra-kin}). Stellar velocity dispersions dispersions were measured using the  CaT lines at
$\lambda\lambda$\,8494, 8542, 8662\,\AA using the cross-correlation technique  described
in detail by \citet{TD79}. Late type giant and
supergiant stars that have strong CaT absorption lines were used 
as stellar velocity templates.

\begin{figure}[tb]
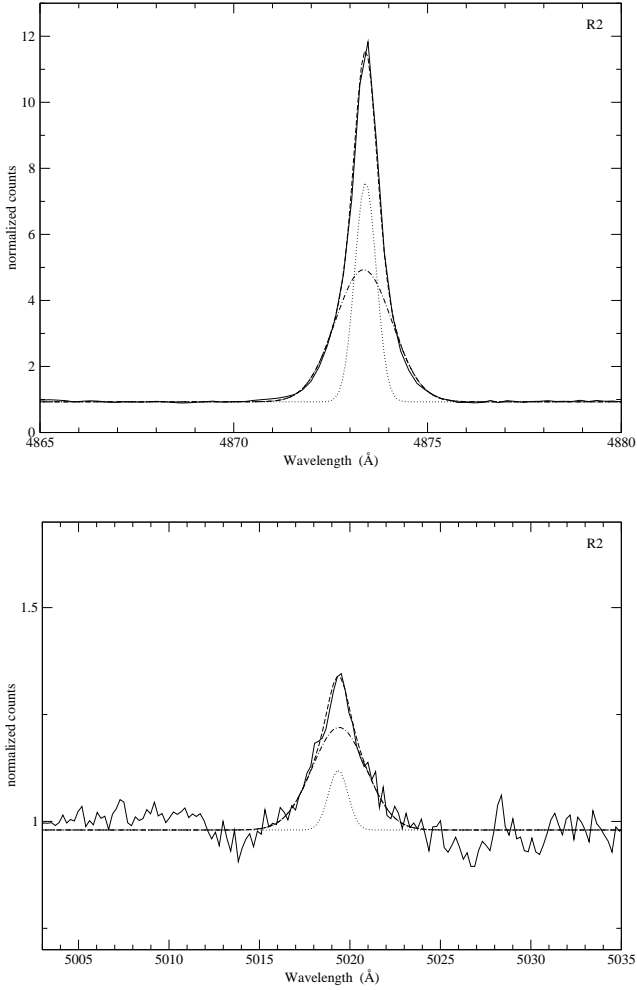

\plotone{ngauss-Hb-R2.eps}
\vspace{0.2cm}
\plotone{ngauss-OIII-R2.eps}
\caption{Sections of the normalised spectrum of region R2 in NGC~3351. Left:  H$\beta$ and right  [O{\sc iii}]\,$\lambda$\,5007\,\AA\ . The dashed-dotted line is the broad component, the dotted line is the narrow component and the dashed line is the sum of both} %% no full stop at the end of caption
\label{velocities}
\end{figure}

For the 5 CNSFR observed in NGC~3351, stellar velocity dispersions are found to be between 39 and 67\,km\,s$^{-1}$, about 20\,km\,s$^{-1}$ larger than those measured for the gas, if a single Gaussian fit is used. However, the best fits obtained involved two different components for the gas: a ``broad
component" with a velocity dispersion similar to that measured for the stars,
and a ``narrow component" with a dispersion lower than the stellar one by
about 30\,km\,s$^{-1}$. These two components were found both in the hydrogen recombination lines (Balmer and Paschen) and in the [OIII] $\lambda$ 5007 \AA\ line. The narrow component is dominant in the H recombination lines, while the broad component dominates the [OIII] one.  Figure \ref{velocities} shows this effect.

CNSFR are seen to consist of several individual star clusters, although some of them seem to have an only knot, at the HST resolution. The derived masses for the individual clusters as derived using the sizes measured on the HST images are between 1.8
and 8.7\,$\times$\,10$^6$\,M$_\odot$. These values are between 5.5 and 26
times the mass derived for the SSC A in NGC\,1569 by
\citet{HF96} and larger than other kinematically derived SSC
masses. 
Values for the dynamical masses of the CNSFRs are in the range between 4.9\,$\times$\,10$^6$ and
4.3\,$\times$\,10$^7$\,M$_\odot$. Masses derived from the H$\beta$ velocity dispersion under the
assumption of a single component for the gas would have been underestimated
by factors between approximately 2 to 4.

The masses of the ionising stellar clusters of the CNSFRs have been derived from
their H$\alpha$ luminosities under the assumption that the regions are
ionisation bound and without taking into account any photon absorption by
dust. Their values are between 8-0\,$\times$\,10$^5$ and
2.5\,$\times$\,10$^6$\,M$_\odot$.  Therefore, the ratio of the ionising
stellar population to the total dynamical mass is between 0.02 and 0.16. 

The SSC in the observed CNSFRs seem to contain composite stellar populations. Although the youngest one dominates the UV light and is responsible for the gas ionisation, it
constitutes only about 10\,\% of the total. This can explain the low
equivalent widths of emission lines measured in these regions.  This may well apply to
the case of other SSC and therefore conclusions drawn from fits of SSP (single
stellar population) models should be taken with caution 
\citep[e.g.][]{MGG03}.
Also the composite
nature of the CNSFRs  means that star formation in the rings is a process that
has taken place over time periods much longer than those implied by the
properties of the ionised gas.

\section{Abundance analysis}
\label{Abundances}

\begin{figure}[tb]
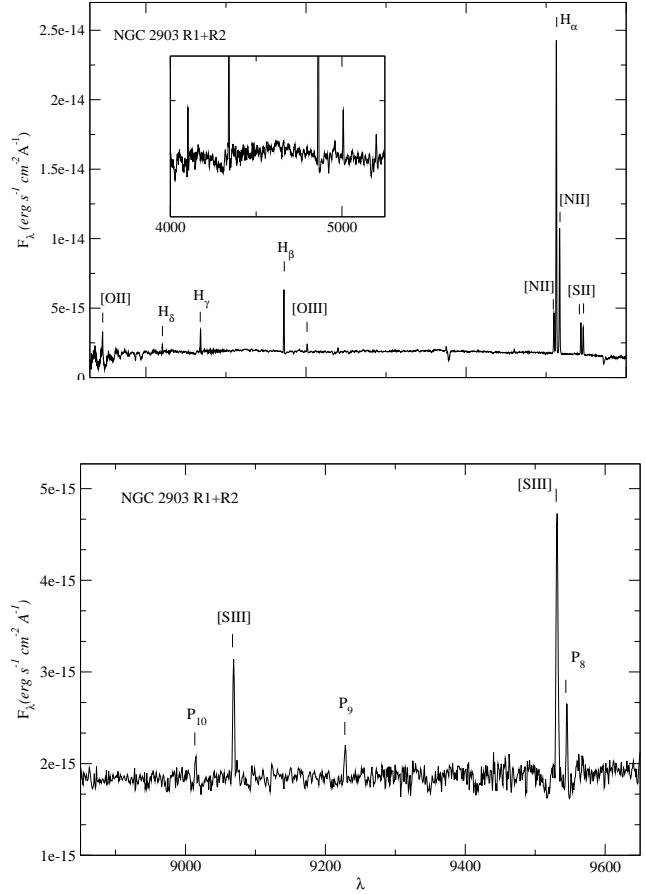

\plotone{2903-r12-blue.eps}
\vspace{0.2cm}
\plotone{2903-r12-red.eps}
%\centering
%\hspace{0.2cm}
%\includegraphics[width=.40\textwidth,height=.25\textwidth,angle=0]{2903-r12-blue.eps}\\
%\vspace{0.9cm}
%\includegraphics[width=.40\textwidth,height=.25\textwidth,angle=0]{2903-r12-red.eps}
\caption{Blue (upper panel) and red (lower panel) 
spectra of region R12 in NGC~2903} 
\label{weaklines}
\end{figure}

For 12 CNSFRs in the galaxies NGC~2903, NGC~3351 and NGC~3504 chemical abundances of O, N and S were derived from the lower resolution spectra. All the spectra are characterized by very weak [OIII] $\lambda\lambda$ 4959, 5007 \AA\ lines. Figure \ref{weaklines} shows the spectrum of a typical region. 

This low excitation is typical of high abundance regions in which the cooling is dominated by [OII] and [OIII] emission lines. In these cases the detection of the auroral [OIII] $\lambda$ 4363 \AA\ line which woukd allow the derivation of the electron temperature is virtually impossible and empirical methods based on the calibration of strong emission lines have to be used. The most commonly used abundance indicator, O$_{23}$, involves the sum of the [OII] and [OIII] emission lines, supposed to be almost independent of geometrical effects in the nebulae. However, this indicator is not well calibrated at the high metallicity end where the oxygen lines are so weak that the value of the R$_{23}$ practically saturates. 

\begin{figure}[tb]
\plotone{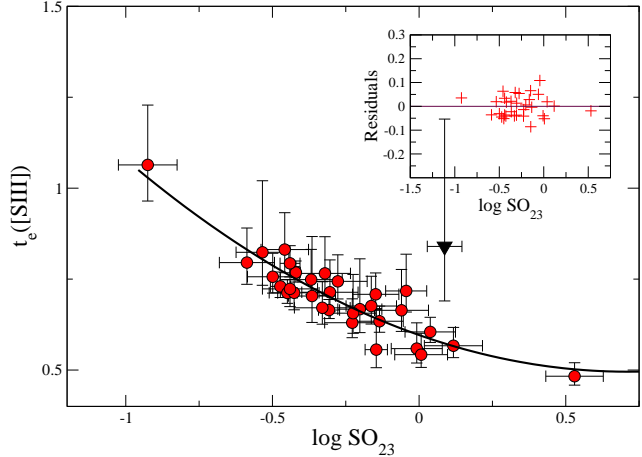}
\caption{Empirical calibration of the [SIII] electron temperature as a function of the abundance parameter SO$_{23}$. The solid line represents a quadratic fit to the high metallicity HII region data } %% no full stop at the end of caption
\label{scalib}
\end{figure}

The sulphur lines, much stronger than the oxygen lines in high metallicity regions, provide an alternative way to derive their abundances.  We have developed a new method for the derivation of sulphur abundances based on the calibration of the [SIII] electron temperature vs the empirical parameter SO$_{23}$ defined as the quotient of the oxygen and sulphur abundance parameters O$_{23}$ and S$_{23}$ and the further assumption that T([SIII]) $ \simeq $ T([SII]). Then the oxygen abundances and the N/O and S/O ratios can also be derived. Figure \ref{scalib} shows this calibration for which data from different sources as listed in \citet{DTCH07} have been used. The actual fit to the data gives:
\[
t_e([SIII]) = 0.596 - 0.283 log SO_{23} + 0.199 (log SO_{23})^2
\]

 Only for one of the regions, the [SIII] $\lambda$ 6312  \AA\ line was detected providing, together with the nebular [SIII] lines at $\lambda\lambda$ 9069, 9532 \AA\ , a value of the electron temperature of T$_e$([SIII])= 8400$^{+ 4650}_{-1250}$K. This value is slightly higher than predicted by the proposed fit and is represented as a solid upside down triangle in Figure \ref{scalib}.

The derived oxygen abundances are comparable to those found in high metallicity disc HII regions from direct measurements of electron temperatures and are consistent with solar values within the errors. The highest oxygen abundance derived is 12+log(O/H) = 8.85, about 1.5 solar \footnote{The solar oxygen abundance is taken as the value derived by \citet{AGS05}, 12+log(O/H)$_{\odot}$ = 8.66$\pm$0.05}. The lowest oxygen abundance derived is about 0.6 times solar.
In all the observed CNSFR the O/H abundance is dominated by the O$^+$/H$^+$ contribution, as is also the case for high metallicity disc HII regions. For our observed regions, however also the  S$^+$/S$^{2+}$ ratio is larger than one. This is not the case for the high metallicity disc HII regions for which, in general, the sulphur abundances are dominated by S$^{2+}$/H$^+$. 
The derived N/O ratios are in average larger than those found in high metallicity disc HII regions and they do not seem to follow the trend of N/O vs O/H which marks the secondary behaviour of nitrogen. On the other hand, the S/O ratios span a very narrow range between 0.6 and 0.8 of the solar value.

\section{Ionized gas characteristics}
\label{Ionization}

From the calculated values of the number of Lyman $\alpha$ photons, Q(H$^0$), ionisation parameter and electron density, it is possible to derive the size of the emitting regions as well as the filling factor \citep[see][]{CDT02}.The derived sizes are between 1.5 arcsec  and 5.7 arcsec which correspond to linear dimensions between 74 and 234 pc. The derived filling factors, between 6 $\times$ 10$^{-4}$ and 1 $\times$ 10$^{-3}$, are lower than commonly found in giant HII regions ($\sim$ 0.01). H$\alpha$ luminosities are larger than the typical ones found for disc HII regions and overlap with those measured in HII galaxies. The region with the largest H$\alpha$ luminosity is R3+R4 in NGC~3504, for which a value of 2.02 $\times$ 10$^{40}$ is measured.  Ionising cluster masses range between  1.1 $\times$ 10$^{5}$ and 4.7 $\times$ 10$^{6}$ M$_{\odot}$ but could be lower by factors between 1.5 and 15 if the contribution by the underlying stellar population is taken into account. These values are consistent with those found from the kinematical data set.

\begin{figure}[tb]
\plotone{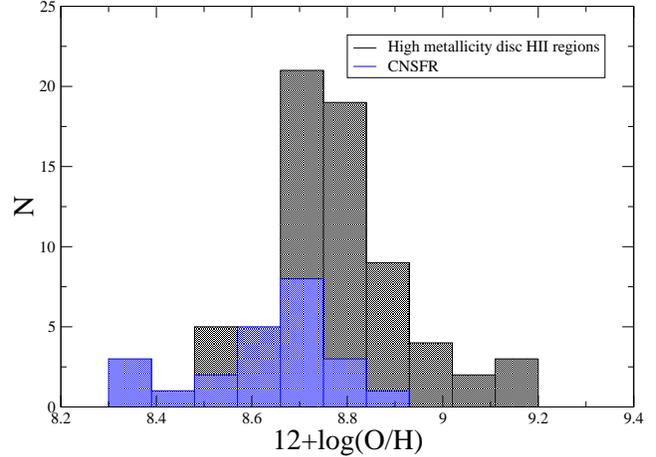}
\vspace{0.6cm}
\plotone{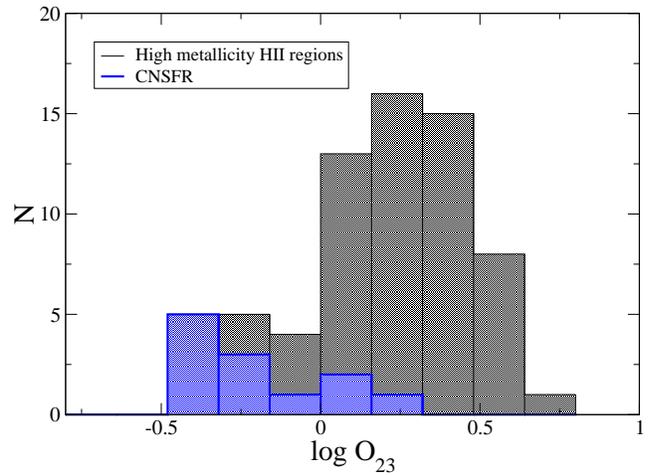}
\caption{Distribution of derived oxygen abundances of CNSFR and high metallicity disc HII regions (top) as compared to that of the O$_{23}$ parameter (bottom). While the first show comparable distributions for the two kinds of objects, the second shows lower O$_{23}$ (higher abundances) for CNSFR} %% no full stop at the end of caption
\label{abun}
\end{figure}

When compared to high metallicity disc HII regions, CNSFR show values of the O$_{23}$ and the N2 parameters whose distributions are shifted to lower and higher values respectively, hence, even though their derived oxygen and sulphur abundances are similar, higher values would in principle be obtained for the CNSFR  if pure empirical methods were used to estimate abundances. This can be seen in Figure \ref{abun} where the distributions of the derived O/H abundances and the O$_{23}$ pararameter for CNSFR and high metallicity disc HII regions are shown.

%\begin{figure}[tb]
%\plotone{histo_S2-S3_2.eps}
%\caption{Distribution of the [SII]/[SIII] ratio for the observed CNSFR (dark) and the sample of high metallicity disc HII regions (light) showing the segregation between the two families of objects in ionisation parameter } %% no full stop at the end of caption
%\label{ionpar}
%\end{figure}

\begin{figure}[tb]
\plotone{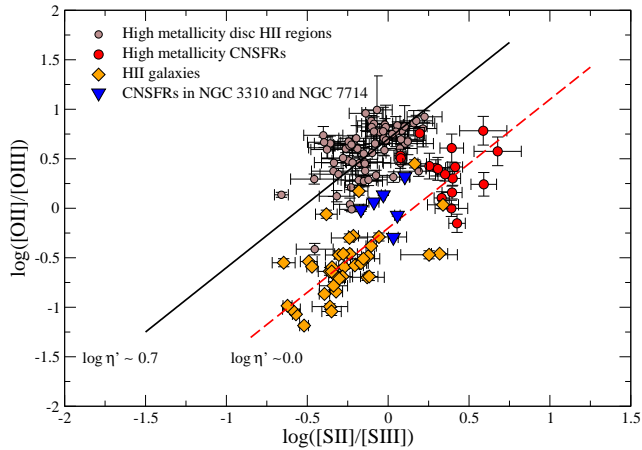}
\caption{Distribution of the [SII]/[SIII] ratio for the observed CNSFR (dark) and the sample of high metallicity disc HII regions (light) showing the segregation between the two families of objects in ionisation parameter } %% no full stop at the end of caption
\label{etaplot}
\end{figure}

CNSFR also show lower ionisation parameters than their disc counterparts, as derived from the [SII]/[SIII] ratio. %(Figure \ref{ionpar}). 
Their ionisation structure also seems to be different with CNSFR showing radiation field properties more similar to HII galaxies than to disc high metallicity HII regions.
This can be seen in Figure \ref{etaplot} that shows  [OII]/[OIII] ratio vs  [SII]/[SIII]. Since the parameter $\eta$, which is the quotient of these two ratios, is a good indicator of ionising temperature \cite{VP88}, diagonal lines in this plot should run along objects with similar temperatures.  CNSFR in this diagram also segregate from the high metallicity disc HII regions pointing to higher values of the ionising temperature.

 The possible contamination of their spectra from hidden low luminosity AGN and/or shocks, as well as the probable presence  of more than one velocity component in the ionised gas corresponding to kinematically distinct systems (see Section 3 above), should be further investigated. 

%% Math 
%
%\begin{eqnarray}%\label{eqn:?}
%\\ \nonumber
%\end{eqnarray}
%
%\begin{equation}%\label{eqn:?}
%\end{equation}

%% Table (two-column)
%
% \begin{table*}%%[tb]
% \small
% \caption{Caption} %% no full stop at the end of caption
%  \label{tbl:?}
% \begin{tabular}{}
% \tableline  %% rule at top
% \tablenotemark{a} 
% <entries>
% \tableline %% rule at bottom
% \end{tabular}
%
%% Any table notes must follow the \end{tabular} command. 
% \tablenotetext{a}{}
% \tablecomments{}
% \end{table*}

%% Table (one-colum)
%
% \begin{table}
% \caption{} %% no full stop at the end of caption
% \label{tbl:?}
% \begin{tabular}{}
% \tableline  %% rule at top
% \tablenotemark{a} 
% <entries>
% \tableline %% rule at bottom
% \end{tabular}
% \end{table}

%% Deluxe table (refer to AASTeX documentation)
%
% \begin{deluxetable}{ccrrrrrrrrcrl}
% \tabletypesize{\scriptsize}
% \rotate
% \tablecaption{}% %% no full stop at the end of caption
% \label{tbl:?} 
% \tablewidth{0pt}
% \tablehead{\colhead{}}
% \startdata
% <entries>
% \enddata
%
%% Text for table notes should follow after the \enddata but before
%% the \end{deluxetable}. Make sure there is at least one \tablenotemark
%% in the table for each \tablenotetext.
% \tablecomments{}
% \tablenotetext{a}{}
% \tablenotetext{b}{}
% \end{deluxetable}

%% Figure 
%
% \begin{figure}%[tb]
% \includegraphics{}
% \includegraphics[width=\columnwidth]{}
% \caption{} %% no full stop at the end of caption
% \label{fig:?}
% \end{figure}

%% Acknowledgements
%
 \acknowledgments
% <Acnowledgments text>
This research has been supported by  DGICYT grant AYA-2004-02860-C03. 

%% References
%% Please cite all reference entries in the article text using \cite or
%% equivalent command. 

%%%  Using BibTeX  (Name-Year style)
%
% \bibliographystyle{spr-mp-nameyear-cnd}  %% BibTeX style
% \bibliography{<bib data>}                %% BibTeX data

%% Non-BibTeX  (Name-Year style)
%
% \begin{thebibliography}{}
% \bibitem[\protect\citeauthoryear{<author>}{<year>]{ref:?}
%    <ref. entry>
% \bibitem[\protect\citeauthoryear{<author>}{<year>]{ref:?}
%    <ref. entry>
% \end{thebibliography}

\end{document}